\documentclass[letterpaper]{jpconf}

\usepackage{graphicx}

\begin{document}

\title{
On Fluctuations of Conserved Charges : Lattice Results Versus 
Hadron Resonance Gas
}

\author{Pasi Huovinen$^1$ and P\'eter Petreczky$^2$ 
}        

\address{$^1$
Institut f\"ur Theoretische Physik,
Johann Wolfgang Goethe-Universit\"at,\\
Max-von-Laue-Stra{\ss}e 1,
60438 Frankfurt am Main, Germany
}

\address{$^2$Physics Department, Brookhaven National Laboratory, 
         Upton, NY 11973, USA}

\begin{abstract}
We compare recent lattice results on fluctuations and correlations
of strangeness, baryon number and electric charge 
obtained with p4 improved staggered action
with the prediction
of hadron resonance gas model. 
We show that hadron resonance gas can describe these fluctuations
reasonably well if the hadron properties are as calculated on the lattice.
\end{abstract}

\section{Introduction}
In recent years a lot of progress has been made in understanding the
QCD transition at non-zero temperature through lattice calculations (see Refs. \cite{detar,petr} for
reviews). In particular, fluctuations of conserved charges have been
studied on the lattice and provided  insight on how the relevant degrees of freedom
change from hadronic to partonic 
\cite{gavaiquenched,Gavai2,milc04,hotqcd,eos005,fodor06,fodor09,fluctuations}.
At low temperatures it is expected that hadrons are good degrees of freedom and
thermodynamic quantities can be well described by hadron resonance gas.
Hadron resonance gas (HRG) turned out to be very successful in
describing particle abundances produced in heavy ion collisions
\cite{PBM}. It was also used to estimate QCD transport coefficients \cite{jaki} 
as well as chemical equilibration rates \cite{johanna} close to the
transition temperature.
Thermodynamic quantities calculated in lattice QCD with
rather large quark mass agree well with the HRG model if the masses of the
hadrons in the model are tuned appropriately to match the large quark
mass used in lattice calculations \cite{redlich1}.  Furthermore, the
ratios of certain fluctuations calculated on the lattice are  also in
 reasonably good agreement with HRG model predictions at
low temperatures \cite{redlich2,redlich3,fluctuations}.

Fluctuations of conserved charges are the most suitable quantities
to test the validity of HRG model. They can be defined as derivatives
of the pressure with respect to chemical potentials and as such appear
in the calculations of thermodynamic quantities at finite baryon density
via Taylor expansion. The recent lattice calculations performed with p4 and asqtad actions at 
or very close to the physical quark mass \cite{hotqcd,eos005,fluctuations} gave fluctuations that
were quite different from the results obtained in HRG model.
It was pointed out that lattice
discretization effects on the hadron spectrum are responsible for
this discrepancy \cite{pasi}. When taking into account the 
lattice spacing dependence of the hadron masses in the hadron resonance
gas calculations a good agreement between lattice and HRG calculations has been
found. Strangeness fluctuations calculated with and stout \cite{fodor06,fodor09} and
HISQ \cite{lat09} improved staggered fermion actions agree better with the HRG result (see 
also the discussion in Ref. \cite{these}). This is due to largely reduced discretization
effects in the hadron spectrum for these actions. In Ref. \cite{pasi} we studied
baryon number and strangeness fluctuations in HRG, where the hadron masses have been modified
to include discretization effects present in lattice calculations. The comparison was done using
lattice data obtained  with asqtad action on $N_{\tau}=6$ and $8$ lattices. 
On the
other hand, the most detailed lattice study of baryon number, strangeness and electric charge fluctuations and
correlations up to sixth order was performed with p4 action using lattices with temporal extent $N_{\tau}=4$
 and $N_{\tau}=6$ \cite{fluctuations}.
In this contribution we are going to extend our calculations 
in modified the HRG model to study different fluctuations and compare them with lattice results obtained 
with p4 action on $N_{\tau}=4$ and $6$ lattices. 
As mentioned above this is also important for constructing
realistic equation of state at finite baryon density for hydrodynamic models, 
similarly as this was done for zero baryon density in Ref. \cite{pasi}.

\section{Fluctuations of strangeness, baryon number and electric charge}
Derivatives of the pressure with respect to chemical potentials of conserved charges,
e.g. baryon number ($B$), electric charge $Q$ and strangeness $S$ at zero chemical potentials
can be well calculated in lattice QCD
\begin{equation}
\chi_n^X=T^n \frac{\partial^n p(T,\mu_B,\mu_Q, \mu_S)}{\partial \mu_X^n}|_{\mu_B=\mu_Q=\mu_S=0},~X=B,Q,S.
\end{equation}
These are related to quadratic and higher order fluctuations
of conserved charges $\chi_2^X=\langle X^2 \rangle/(V T^3)$,
$\chi_4^X=( \langle N_X^4 \rangle - 3 \langle N_X^2 \rangle^2)/(VT^3)$
etc.\footnote{Here we consider the case of zero chemical potential, so $\langle N_X \rangle=0$.}
Mixed derivatives of the pressure give correlations of conserved charges
\begin{equation}
\chi_{11}^{XY}=T^2 \frac{\partial^2 p(T,\mu_B,\mu_Q, \mu_S)}{\partial \mu_X \partial \mu_Y}|_{\mu_B=\mu_Q=\mu_S=0}=
\langle X Y \rangle/(V T^3),~X,Y=B,Q,S.
\end{equation}
We have calculated these quantities in HRG with hadron spectrum modified to take into account
the lattice artifacts. Unfortunately, there is no detailed calculation of the hadron spectrum
with p4 action. Therefore the lattice spacing and
quark mass dependence of the hadron masses was evaluated using the formulas derived in Ref. \cite{pasi}
based on asqtad calculations. We expect that cutoff dependence of the hadron masses is similar for
p4 and asqtad actions, except in the pseudo-scalar meson sector. It is known that the quadratic
splitting of non-Goldstone pseudo-scalar mesons is about two times larger for the p4 action than for
the asqtad action \cite{scalingtest}. Therefore when evaluating the contribution of kaons and pions
to different quantities we simply doubled the quadratic splittings in the pseudo-scalar meson 
sector used in Ref. \cite{pasi}. 
To get agreement between lattice results and HRG it turned out to be necessary to modify 
the masses of excited baryons states. Since the cutoff dependence of the excited baryon masses
is not known, it was assumed that masses of all excited states up to certain threshold $m_{cut}^B$
are modified the same way as the ground state baryon masses \cite{pasi}. Values $m_{cut}^B=(1.7-2.5)$GeV
have been considered in the previous analysis. 

\begin{figure}
\includegraphics[width=0.45\textwidth]{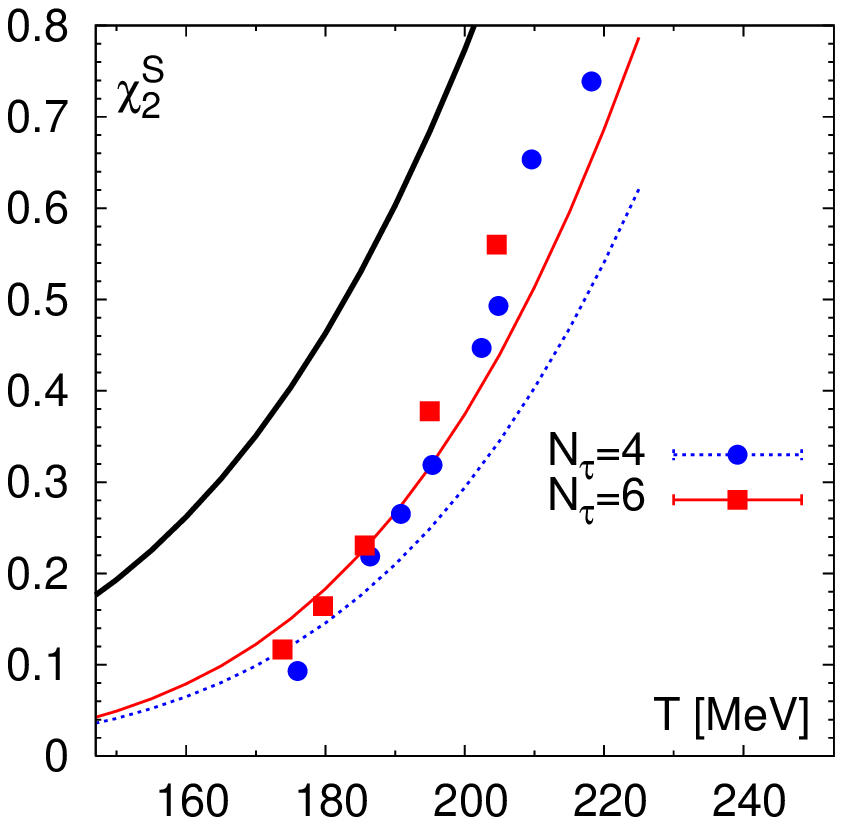}
\includegraphics[width=0.45\textwidth]{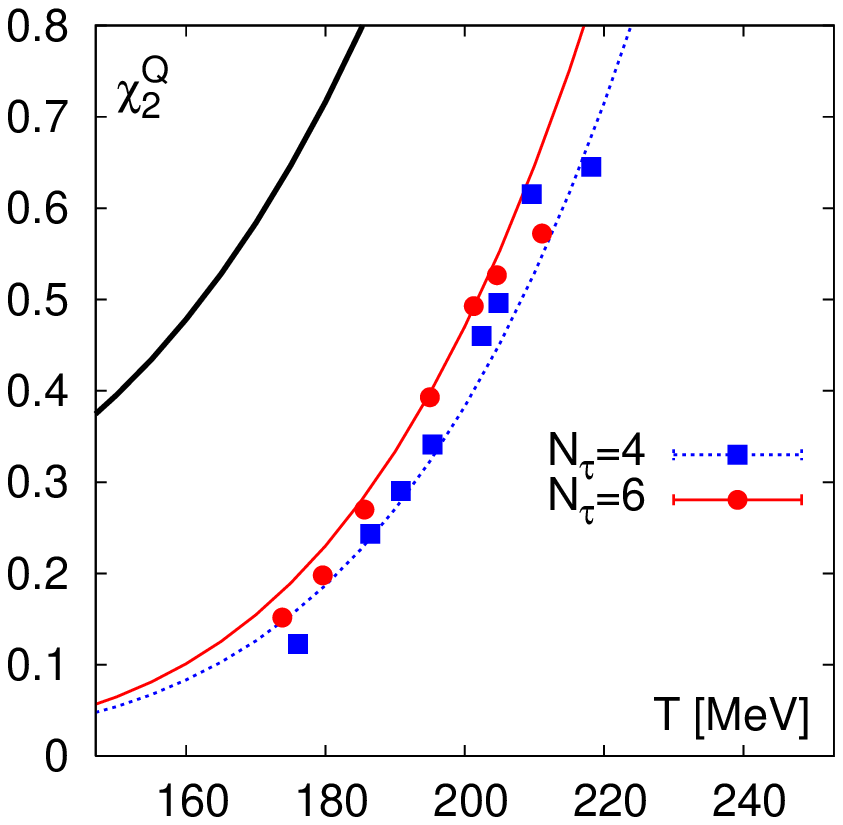}
\vspace*{-0.2cm}
\caption{The quadratic fluctuations of strangeness (left) and electric charge (right)
calculated on the lattice with p4 action \cite{fluctuations} and compared with the HRG calculations in
the continuum (solid black line) and on the lattice (blue and red lines).}
\label{fig:chi2}
\vspace*{-0.2cm}
\end{figure}

\begin{figure}
\includegraphics[width=0.45\textwidth]{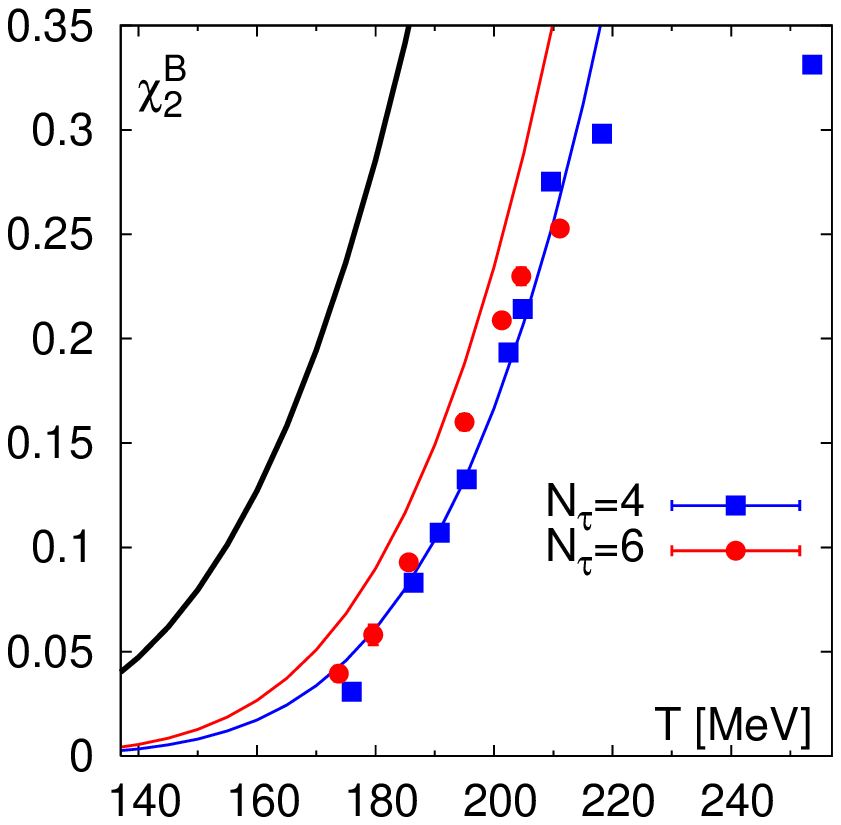}
\includegraphics[width=0.45\textwidth]{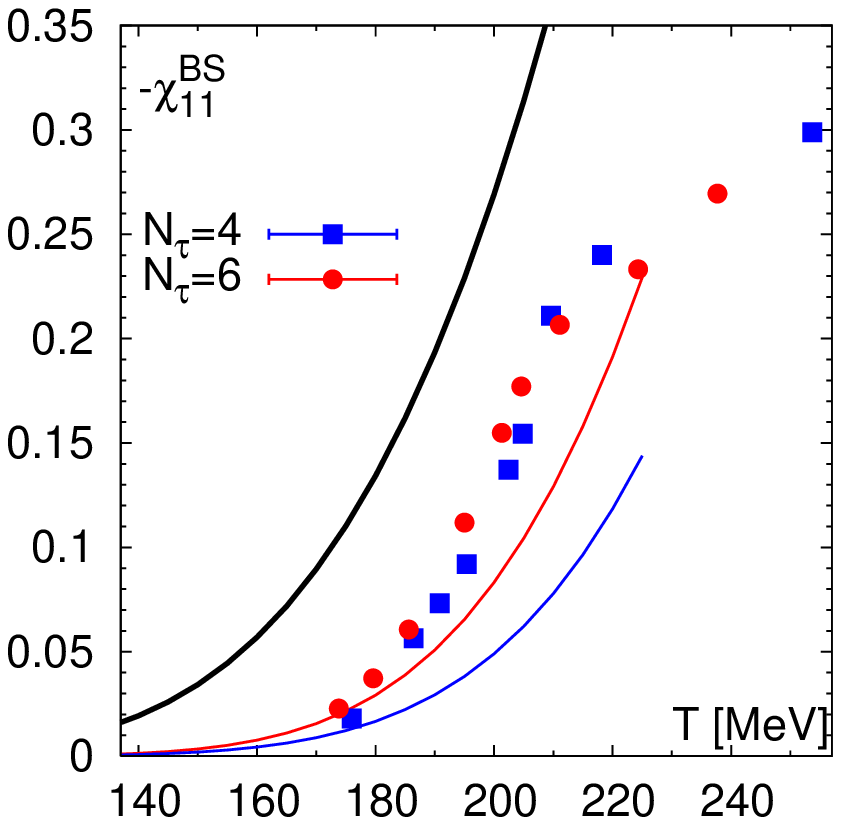}
\vspace*{-0.2cm}
\caption{The quadratic fluctuations of baryon number (left) and baryon number - strangeness
correlations (right)
calculated on the lattice with p4 action \cite{fluctuations} and compared with the HRG calculations in
the continuum (solid black line) and on the lattice (blue and red lines).}
\label{fig:chiB}
\vspace*{-0.2cm}
\end{figure}

\begin{figure}
\includegraphics[width=0.45\textwidth]{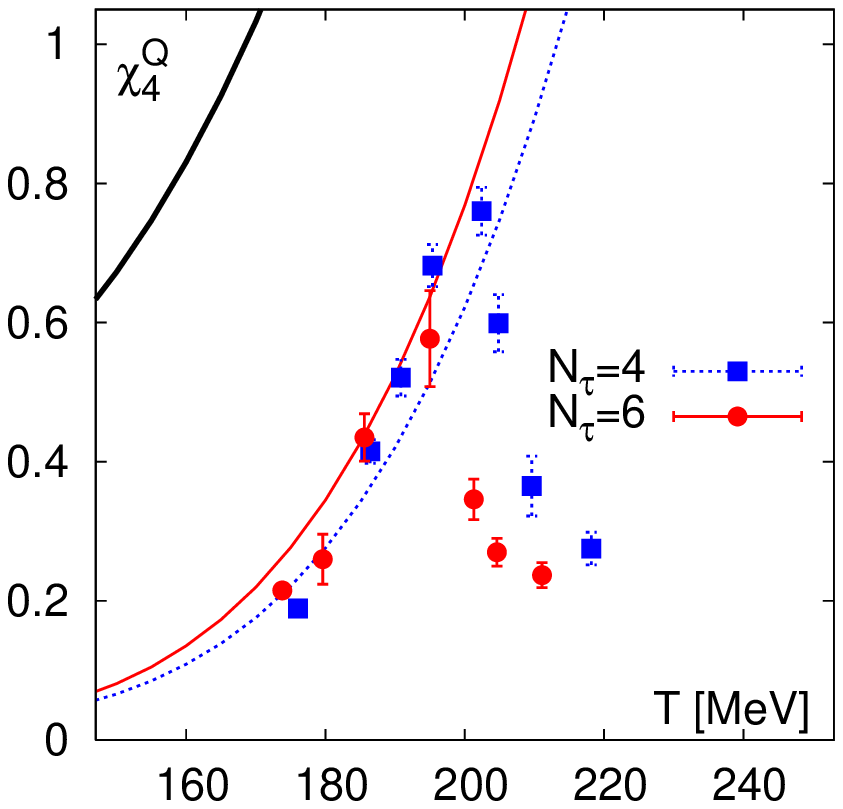}
\includegraphics[width=0.45\textwidth]{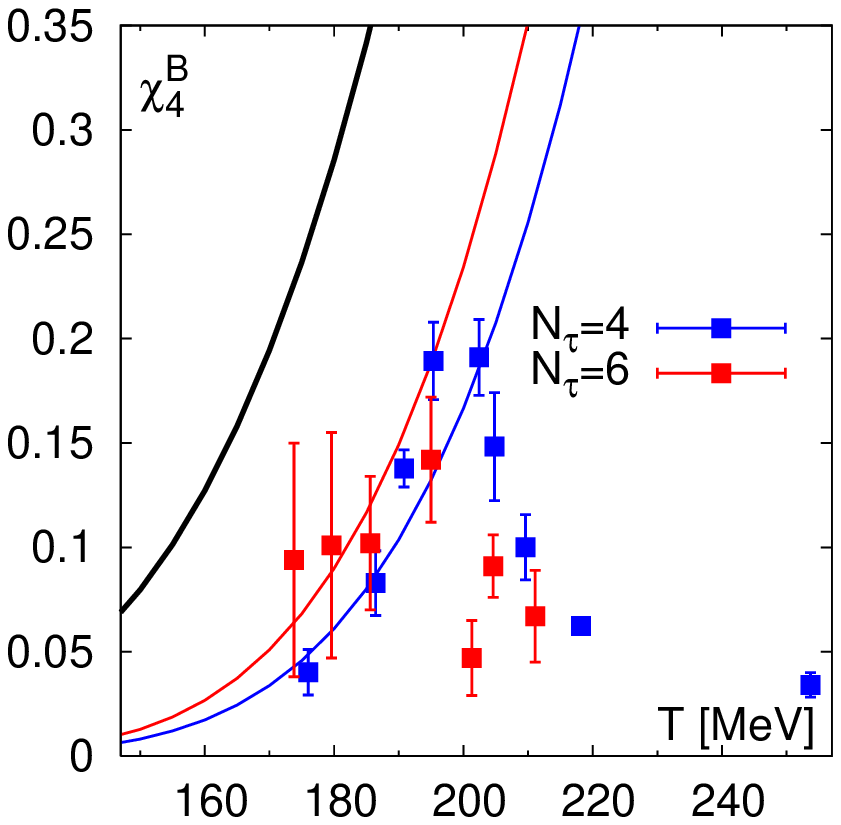}
\vspace*{-0.2cm}
\caption{The fourth order fluctuations of the electric charge (left) and baryon number (right)
calculated on the lattice with p4 action \cite{fluctuations} and compared with the HRG calculations in
the continuum (solid black line) and on the lattice (blue and red lines).}
\vspace*{-0.2cm}
\label{fig:chi4}
\end{figure}

In this paper we consider all resonances with mass up to $2.5$GeV and use $m_{cut}^B=1.9$GeV
in all calculations.
The quadratic fluctuations of strangeness and electric charge calculated in HRG and compared
to the p4 lattice results are shown in Fig. \ref{fig:chi2}. The lattice results are well below the HRG
curve obtained with physical hadron masses. This is a general feature for all fluctuations.
However, taking into account the discretization effects
in the hadron spectrum we find a reasonable agreement between HRG and lattice results. In the strange sector
discretization effects are slightly overestimated in our approach. We also studied quadratic 
baryon number fluctuations and baryon number - strangeness correlations and the corresponding
results are shown in Fig. \ref{fig:chiB}. The baryon number fluctuations are well described by modified
HRG, although the in the low temperature region cutoff effects are under-predicted for $N_{\tau}=6$.
For baryon number-strangeness fluctuations the agreement between lattice and modified HRG is not very good. 
This may imply that the cutoff effects in strange baryon sector are smaller than in non-strange baryon
sector and are overestimated in our approach. Interestingly, the new calculations with HISQ action also
suggest smaller cutoff effects in baryon number-strangeness fluctuations compared to baryon number
fluctuations \cite{these}. Finally we have considered fourth order fluctuations of the electric charge
and baryon number. The numerical results are shown in Fig. \ref{fig:chi4}. The HRG calculations
with the modified hadron masses agree reasonably well with the p4 lattice data. Note, however,
that here the deviations from the resonance gas show up at smaller temperatures. 

\section{Conclusions}
We have studied quadratic and fourth order fluctuations of baryon number, electric charge
and strangeness fluctuations in HRG model, where hadron masses have been adjusted to take
into account the discretization errors in the hadron spectrum present in the lattice calculations. We have found
reasonably good agreement with the lattice calculations of the fluctuations performed using p4 action on $N_{\tau}=4$
and $6$ lattices. Our calculations explain why all the lattice results fall below the physical HRG
result. Our approach does not give a good description of baryon number - strangeness fluctuations.
To resolved this issue a  more refined treatment of the cutoff effects in the baryon sector is needed.

\section*{Acknowledgments}
This work was supported by the U.S. Department of Energy under
contract DE-AC02-98CH1086 and by the ExtreMe Matter Institute
(EMMI). P.H.~is grateful for support from Center of Analysis and
Theory for Heavy Ion Experiment (CATHIE) which enabled him to stay in
BNL where large part of this work was finalized.

\end{document}